\documentclass[conference]{IEEEtran}

\usepackage[utf8]{inputenc}

\usepackage{csquotes}

\usepackage[numbers]{natbib}

\usepackage{graphicx}
\graphicspath{{images/}}

\usepackage{amsmath}

\PassOptionsToPackage{hyphens}{url}\usepackage{hyperref}

\usepackage{array}

\usepackage{multirow}

\usepackage{xcolor}

\definecolor{orange1}{RGB}{233, 194, 175}
\definecolor{orange2}{RGB}{215, 125, 102}
\definecolor{orange3}{RGB}{166, 57, 55}

\definecolor{blue1}{RGB}{206, 219, 232}
\definecolor{blue2}{RGB}{131, 169, 197}
\definecolor{blue3}{RGB}{67, 110, 158}

\definecolor{green1}{RGB}{207, 227, 199}
\definecolor{green2}{RGB}{138, 184, 132}
\definecolor{green3}{RGB}{69, 123, 78}

\usepackage{float}

\usepackage{tcolorbox}

\def\BibTeX{{\rm B\kern-.05em{\sc i\kern-.025em b}\kern-.08em
    T\kern-.1667em\lower.7ex\hbox{E}\kern-.125emX}}

\begin{document}

\title{On the Effectiveness of Microservices Tactics and Patterns to Reduce Energy Consumption: An Experimental Study on Trade-Offs}

\author{
    \IEEEauthorblockN{Xingwen Xiao}
    \IEEEauthorblockA{
        Vrije Universiteit Amsterdam\\
        Amsterdam, The Netherlands\\
        im.xingwen.xiao@gmail.com
    }
    \and
    \IEEEauthorblockN{Chushu Gao}
    \IEEEauthorblockA{
        Software Improvement Group\\
        Amsterdam, The Netherlands\\
        chushu.gao@softwareimprovementgroup.com
    }
    \and
    \IEEEauthorblockN{Justus Bogner}
    \IEEEauthorblockA{
        Vrije Universiteit Amsterdam\\
        Amsterdam, The Netherlands\\
        j.bogner@vu.nl
    }
}


\maketitle


\begin{abstract}

\textit{Context:} Microservice-based systems have established themselves in the software industry.
However, sustainability-related legislation and the growing costs of energy-hungry software increase the importance of energy efficiency for these systems.
While some proposals for architectural tactics and patterns exist, their effectiveness as well as potential trade-offs on other quality attributes (QAs) remain unclear.

\textit{Goal:} We therefore aim to study the effectiveness of microservices tactics and patterns to reduce energy consumption, as well as potential trade-offs with performance and maintainability.

\textit{Method:} Using the open-source Online Boutique system, we conducted a controlled experiment with three tactics (\textit{Distribute pods to different nodes}, \textit{Distribute microservices to different containers}, \textit{Use different proxies for different demands}) and three patterns (\textit{Backends for Frontends}, \textit{Request Bundle}, \textit{Caching of Read Requests}) and analyzed the impact of each technique compared to a baseline.
We also tested with three levels of simulated request loads (low, medium, high).

\textit{Results:} Request load moderated the effectiveness of reducing energy consumption.
All techniques (tactics and patterns) reduced the energy consumption for at least one load level, up to 5.6\%.
For performance, the techniques could negatively impact response time by increasing it by up to 25.9\%, while some also decreased it by up to 72.5\%.
Two techniques increased the throughput, by 1.9\% and 34.0\%.
For maintainability, three techniques had a negative, one a positive, and two no impact.

\textit{Conclusion:} Some techniques reduced energy consumption while also improving performance.
However, these techniques usually involved a trade-off in maintainability, e.g., via more code duplication and module coupling.
Overall, all techniques significantly reduced energy consumption at higher loads, but most of them sacrificed one of the other QAs.
This highlights that the real challenge is not simply reducing energy consumption of microservices, but to achieve \textit{energy efficiency}.
\end{abstract}

\begin{IEEEkeywords}
microservices, tactics, patterns, energy consumption, performance, maintainability, controlled experiment
\end{IEEEkeywords}

\section{Introduction}
Over the last few years, microservices have more and more replaced traditional monolithic systems for cloud-based applications.
A microservice-based system decomposes its functionality into smaller, independent services that are connected through lightweight communication protocols~\cite{alshuqayran2016systematic}.
Many companies like Amazon and Netflix have adopted microservices in their cloud infrastructures~\cite{thones2015microservices}.
Typical advantages of microservices are improved maintainability via firm module boundaries, ease of deployment, and scalability~\cite{10220070}.

One less frequently discussed quality attribute (QA) for microservices is environmental sustainability, with energy efficiency as the most important sub-characteristics.
The growing concern for sustainable software is a global phenomenon, and industry must determine how to best move towards green software and green IT~\cite{Verdecchia2021}.
However, there is currently a substantial gap between microservices-related research on environmental sustainability and industrial needs~\cite{araujo2024energy}.
Similar to the whole ICT industry, microservices practitioners are pressured to improve environmental sustainability due to three reasons.
First, they (will) have to comply with new or incoming legislation regarding sustainability~\cite{GSFLegislation2023}.
Second, environmentally unsustainable software often increases business spending on operations and maintenance, such as data center cooling.
Many clusters are so power-consuming that they are put in regions with cheap electricity, i.e., it is also economically advantageous to enhance the energy efficiency of microservice-based systems.
Lastly, the ICT industry contributes to environmental pollution via greenhouse gas emissions due to its energy demand~\cite{Fonseca2019}, which both employees and customers can perceive negatively.

Typical techniques to improve QAs are architectural tactics and patterns.
\citet{bachmann2003deriving} explain architectural tactics as strategic approaches or methods to design software systems to achieve specific quality goals.
This involves making design decisions to directly influence system-level QAs.
In our case, the main quality goal is energy consumption.
\citet{osses2018exploratory} define architectural patterns as abstract descriptions of a system's structure and the behavior within that structure.
Patterns provide a blueprint for organizing the components of a system in a way that addresses particular recurring problems.
In practice, it is also essential to be aware of potential trade-offs with other QAs when applying tactics and patterns.

While some techniques to improve the energy efficiency of microservices have started to emerge~\cite{araujo2024energy}, empirical evidence for their effectiveness is still unclear.
Tactics and patterns are often documented without knowing their concrete effects and downsides~\cite{Vale2022}, e.g., without conducting experiments.
Sometimes, evidence from controlled environments about the impact of such techniques also does not carry over to industrial contexts~\cite{funke2024experimental}.
Having reliable empirical evidence for the effectiveness of such energy techniques, as well as potential trade-offs with other QAs, is very important for microservices practitioners trying to improve environmental sustainability.

Therefore, our goal is to analyze the impact of tactics and patterns for microservices regarding energy consumption and trade-offs with maintainability and performance, two other important QAs in microservice-based systems.
We conducted a controlled experiment with three tactics and three patterns to compare measurements to an unoptimized baseline.
Our results can guide microservices practitioners in selecting suitable techniques, while navigating potential trade-offs.

\section{Background and Related Work}
\label{s:background}

In this section, we present related work about microservices tactics and patterns, as well as their energy efficiency. Microservices represent an architectural style that emphasizes encapsulating business-specific functionality within fine-granular services that communicate with lightweight mechanisms~\cite{Newman2015}.
The single responsibility of each service enables independent testing, deployment, and scaling~\cite{thones2015microservices}.
Due to its widespread adoption in industry, microservices architecture has attracted significant attention in academic research.

Two influential concepts when designing microservice-based systems are \textbf{architectural tactics and patterns}, and several literature studies try to provide an overview.
\citet{osses2018exploratory} conducted a systematic literature review, identifying 44 patterns and tactics for microservices across 8 papers.
In their review, they proposed a taxonomy of microservices patterns with five categories.
While they discussed the QAs associated with the found patterns, energy consumption or efficiency were not mentioned.
Similarly, \citet{Valdivia2019} conducted a systematic literature review about the relationship between 34 microservices patterns and 6 QAs, again without energy-related considerations.
They reported that clearly defined relationships are often lacking, highlighting the need for more empirical research.
In a third systematic literature review, \citet{Li2021} focused on microservices tactics and the associated improved QAs.
They identified 19 tactics that address 6  QAs in microservices architecture, again neglecting energy consumption.
While these reviews summarize the available microservices tactics and patterns, the detailed QA impact of applying these techniques is still an understudied area.
\citet{Vale2022} tried to alleviate this via an interview study with 9 microservices practitioners.
They synthesized a mapping of 14 patterns and their perceived positive and negative QA impact.
Quantitative empirical evidence has also started to accumulate in the area of evolvability~\cite{Bogner2019-PeerJ}, performance~\cite{Pinciroli2023}, reliability~\cite{ElMalki2023}, and understandability~\cite{Bogner2024}.
Nonetheless, the energy consumption impact of microservices patterns and tactics remains an open question.

However, the \textbf{energy consumption of microservices} has been studied from other perspectives than tactics and patterns, with a recent review by \citet{araujo2024energy} curating 37 primary studies.
While this is not a large number and not all of these papers directly focused on energy consumption, it is reasonable to assume that the environmental sustainability of microservices will become an increasingly prominent quality concern among practitioners.
To date, it has still received comparatively limited attention in research and practice.
One example is a proactive scheduling model proposed by \citet{FontanadeNardin2021} to reduce energy consumption with minimal impact on performance through resource elasticity heuristics.
Their evaluation with a microservice-based application in a private cloud achieved a 27.9\% reduction in energy consumption.
Similarly, \citet{Berry2024} investigated the effect of resource elasticity by comparing horizontal scaling of a monolithic application to its microservices version.
The microservices version outperformed the monolithic one in energy use when configured for optimal availability and performance.
Lastly, \citet{Cortellessa2024} applied a genetic algorithm to generate architectural deployment alternatives of microservices with optimal trade-offs between performance, deployment cost, and power consumption.
They found a significant negative impact on response time when a power consumption objective was introduced.
Overall, broad empirical evidence on the effectiveness and trade-offs of energy techniques for microservices is unavailable.

\section{Study Design}
\label{s:design}
To start the accumulation of evidence to fill this gap, we formed an academia-industry collaboration between VU Amsterdam and the Software Improvement Group (SIG)\footnote{\url{https://www.softwareimprovementgroup.com}}, a software consultancy firm specialized in software quality with roughly 160 employees.
Together, we first analyzed the microservices literature to identify suitable tactics and patterns to evaluate.
Afterward, we conducted a controlled experiment~\cite{Wohlin2024}.
In this section, we describe the design of this experiment.
For transparency and reproducibility, we also published our experiment artifacts on Zenodo.\footnote{\url{https://doi.org/10.5281/zenodo.12730144}}

Our research aims to find out how effective existing tactics and patterns are to reduce energy consumption in microservice-based systems and if any trade-offs with other important quality attributes exist.
This is specified more clearly in the following research questions:

\begin{itemize}
    \item[\textbf{{RQ}1}] How effective are proposed architectural tactics and patterns to reduce the energy consumption of microservice-based systems?
    \item[\textbf{{RQ}2}] How does applying these tactics and patterns impact performance and maintainability?
\end{itemize}

To keep the scope for the trade-off analysis manageable (RQ2), we focused on two quality attributes that are generally seen as drivers for using a microservice architecture, performance and maintainability~\cite{Bogner2019}.
A second reason was that performance and energy consumption are sometimes considered as a trade-off~\cite{Cortellessa2024}.

\subsection{Experiment Objects}
\label{sec:Experiment Objects}
Research on architectural tactics and patterns for improving the environmental sustainability of microservices is unfortunately still scarce.
Choosing promising candidates as experiment objects based on our review of the literature was therefore not easy.
As a result, we also included tactics and patterns that have not been specifically conceptualized in the context of energy efficiency, but seemed promising to reduce energy consumption nonetheless.
To guide our selection of candidates, we applied the following inclusion criteria:

\textbf{Not tied to a specific cloud provider}: If a candidate relies too strongly on proprietary cloud infrastructures and tools from AWS, Azure, or Google Cloud, it is difficult to generalize it to other platforms, and also not locally implementable.

\textbf{Ease of implementation}: the candidate can be implemented in our experimental system with reasonable effort.

\textbf{Potentially energy-saving}: the candidate has been either directly recommended to lower energy consumption or it seems reasonable to assume energy-related benefits, e.g., due to reduced computing resource usage.

\textbf{Commonly used}: for candidates that were not directly recommended for energy efficiency, we additionally ensured that they are reasonably popular, e.g., in cloud provider documentation or industrial projects.

To set a manageable scope for the experiment, we selected six candidates, three tactics (T1-T3) and three patterns (P1-P3).
We briefly introduce them below.

\textbf{T1 -- Distribute pods to different nodes to take advantage of node properties:} nodes vary in properties such as energy source, specialized hardware, and workloads from time to time.
These differences are crucial when strategically distributing pods across nodes for optimal load or energy balancing~\cite{valera2022pisco}.
We tested this tactic from the load-balancing perspective, but not from the energy-balancing perspective, as the nodes of our experiment server are based on the same electricity grid.
The number of nodes remained unchanged compared to the control group.
We used the Kubernetes Scheduler\footnote{\url{https://kubernetes.io/docs/concepts/scheduling-eviction/kube-scheduler}} to dynamically assign pods to nodes based on their resource requirements and dependency relationship.

\textbf{T2 -- Distribute microservices to different containers based on affinity:} this energy tactic uses containerized grouping to place several microservices in one container, e.g., services dependent on each other or using the same tech stack~\cite{ContainerizedGrouping}.
We implemented grouping based on programming languages and resolved library dependencies accordingly.

\textbf{T3 -- Use different proxies for different demands}: this tactic proposes to use the most efficient proxy that is suitable for the workload of the system~\cite{IstioImpact}, which should reduce resource usage.
We tested this tactic with Istio\footnote{\url{https://istio.io/latest/about/service-mesh}} and kube proxy\footnote{\url{https://kubernetes.io/docs/reference/command-line-tools-reference/kube-proxy}}, two of the most popular microservices proxies, which are suitable for different workloads~\cite{IstioImpact}.
The experiment baseline used kube proxy and the treatment Istio.

\textbf{P1 -- Backends for Frontends (BFF)}: this pattern involves creating separate orchestration services tailored to the needs of individual clients or user interfaces~\cite{BackendsForFrontends}.
This allows optimizing each backend's functionality, performance, and development process without impacting other client experiences.
It can decrease, e.g., the resources sent to the frontend, the styling description files, and the unused functionality programs on some frontends, making it potentially energy-saving~\cite{Rani2024}.
We implemented two BFFs in the treatment version: one for mobile clients and one for desktop clients.
The mobile frontend is more straightforward in terms of interface, which means that fewer resources will be prepared by the corresponding backend.
Each frontend received one half of the requests.

\textbf{P2 -- Request Bundle:} this pattern describes a batched API request approach to bundle multiple small individual API requests together~\cite{Zimmermann2022}.
It aims to reduce network congestion and hardware resource utilization.
The reduced number of requests and responses may also simultaneously reduce the energy consumption~\cite{Rani2024}.
We implemented request bundles for API endpoints with three responses at most, with the system waiting for three requests at a time.

\textbf{P3 -- Caching of Read Requests:} caching is a frequently used software performance technique that can also be applied to microservices~\cite{Gottlieb2023}.
If requests often read similar data, caching can eliminate the need to communicate with other services or databases.
While it aims to mitigate network congestion and improve API throughput, the Green Software Foundation also lists it as a green software pattern~\cite{GreenSoftwareFoundation_Cache_Static_Data}.
We implemented a simple in-memory cache per service that stored recent requests, without adding a third-party tool like Redis.

\subsection{Experimental Materials}
\label{sec:Experimental Materials}
We required several materials for this experiment.
First, we needed to select a \textbf{suitable microservice-based system}.
For reproducibility, we decided not to use a proprietary system from SIG but instead an open-source system.
A wide variety of such microservice-based systems using different programming languages are available.\footnote{See, e.g., \url{https://github.com/mfornos/awesome-microservices}}
We wanted a system that is actively maintained (at least one commit in the past six months, without a large number of old unresolved bug reports), popular (a sizeable number of GitHub stars), locally executable (not tied to a cloud provider), technologically diverse (not only one programming language), and of manageable size (somewhere between 5 and 20 services).
Online Boutique\footnote{\url{https://github.com/GoogleCloudPlatform/microservices-demo}} is a microservice-based example system from Google Cloud that fulfills these criteria.
It certainly is popular (over 16k stars), and it consists of 10 services in 5 different programming languages (see Table~\ref{table:service_language}).

\begin{table}[ht]
    \centering
    \caption{Services of the Online Boutique System}
    \begin{tabular}{ll}
    \textbf{Service}            & \textbf{Language} \\
    \hline
    \hline
    frontend                    & Go                \\
    cartservice                 & C\#               \\
    productcatalogservice       & Go                \\
    currencyservice             & Node.js           \\
    paymentservice              & Node.js           \\
    shippingservice             & Go                \\
    emailservice                & Python            \\
    checkoutservice             & Go                \\
    recommendationservice       & Python            \\
    adservice                   & Java              \\
    \hline
    \hline
    \end{tabular}
    \label{table:service_language}
\end{table}

To execute the experiment, we used an \textbf{experiment server cluster} in VU Amsterdam's Green Lab~\cite{procaccianti_green_2015,mendez_ten_2024}, which is specifically designed for energy efficiency experiments.
Using this local environment enabled us to collect energy consumption readings and remove unknown infrastructure configurations.
The used machines had an Intel Xeon Silver 4208 CPU @ 2.10 GHz and 395 GB of RAM.
Intel CPUs offer Running Average Power Limit (RAPL), which is a feature that allows to fetch accurate energy consumption readings from CPUs and DRAM~\cite{RAPL}.
Many more advanced tools are built on top of RAPL.
For our microservices experiment, we chose Scaphandre\footnote{\url{https://github.com/hubblo-org/scaphandre}}, which integrates well with Kubernetes and the Grafana and Prometheus monitoring stack.
This also allowed generating intuitive graphs to compare different system versions.

Moreover, we needed a \textbf{load generator} to simulate API requests during the experiment.
Several popular tools are available for this, but we needed a non-commercial generator that also allows for customized test functionality via scripting.
We therefore selected Locust\footnote{\url{https://locust.io}}, which is also the default load generator that can be enabled in Online Boutique, thereby lowering the barrier for troubleshooting the experiment environment.
Locust test cases can be scripted using Python, making it highly flexible and customizable.

\begin{figure}[ht]
    \centering
    \includegraphics[width=0.85\linewidth]{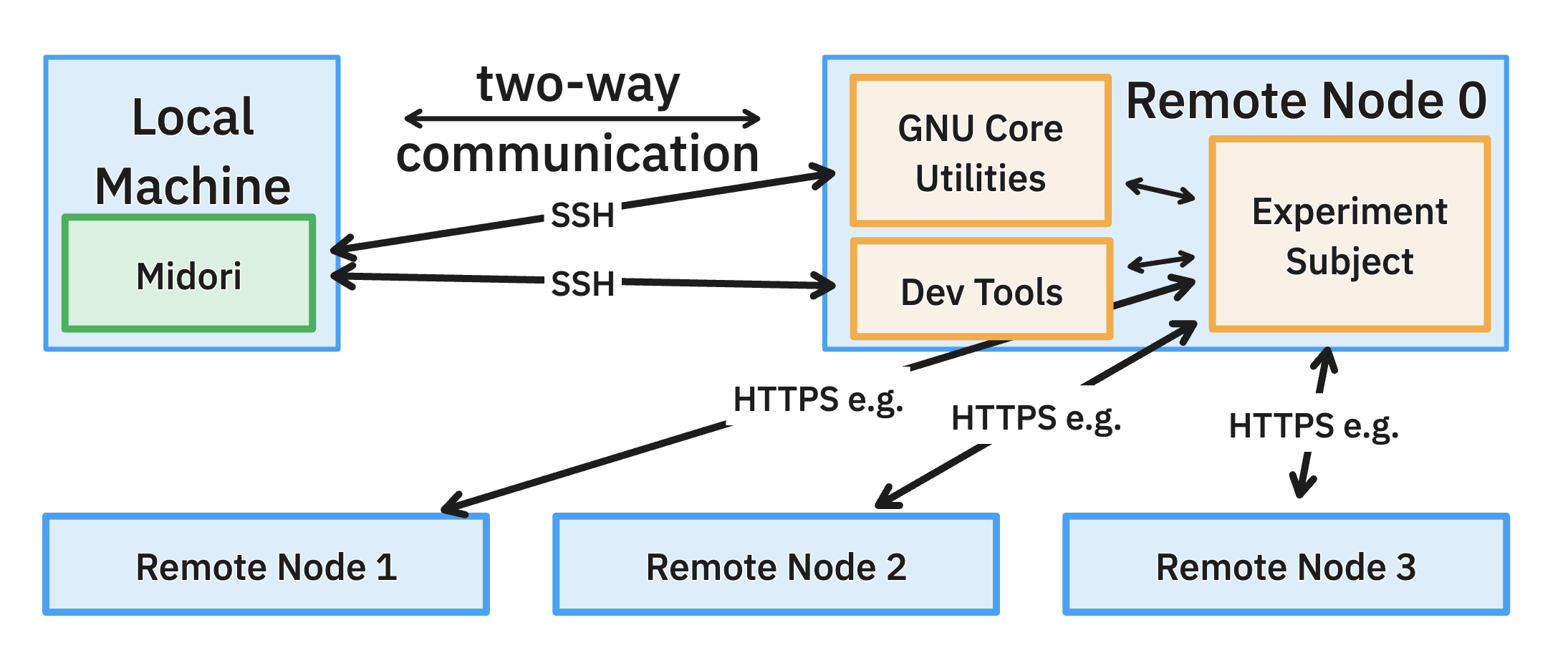}
    \caption{Experiment Environment and Orchestrator}
    \label{fig:orchestrator architecture model}
\end{figure}

Due to the large number of experiment trials, we also needed an \textbf{experiment orchestrator} for automation.
Since we were not satisfied with existing offerings and SIG had wanted to implement a custom orchestrator for some time, we developed Midori\footnote{\url{https://github.com/cyanxiao/midori}}, a plugin-based tool written in Python.
Midori simplifies recurring tasks during experiments, like establishing and closing SSH connection, randomizing the order of trial sequences, adhering to a cooldown period, switching between different treatments, etc.
It interacts with the experiment environment as shown in Fig.~\ref{fig:orchestrator architecture model}.
Via the main remote node, Midori can either access GNU core utilities like \texttt{ls}, \texttt{cd}, \texttt{ps}, etc., or development tools like \texttt{top}, \texttt{kubectl}, etc.
These two kinds of executables are sufficient to manipulate the experiment environment and to collect data. 

For RQ2, we also needed a \textbf{maintainability analysis tool}.
SIG offers a code quality analysis tool named \textit{Sigrid}\footnote{\url{https://docs.sigrid-says.com}}, which is based on ISO~25010:2011.
One of its core functions is a system-level maintainability analysis, which includes maintainability sub-variables like duplication of code, unit size, unit complexity, unit interfacing complexity, and module coupling.
The results of maintainability sub-variables are shown as scores ranging between $0.5$ and $5.5$.
These scores are based on more than 10,000 software systems previously benchmarked by the tool, where the lower the score, the worse it compares to other software systems regarding this sub variable.
The higher the score, the better.
For example, the unit interfacing score is higher when the complexity of unit interfacing is down.
A score of $3.0$ in the duplication of codes means that half of the benchmarked systems perform worse than the current one in terms of the duplication of codes.

\subsection{Experiment Variables}
\label{sec-Variables}
The most important \textbf{independent variable} for our experiment was the applied tactic or pattern.
In addition to the baseline without any modifications, this included the discussed three tactics (T1-T3) and three patterns (P1-P3), i.e., this variable had a total of seven levels.
The second consciously manipulated independent variable was the load we imposed on the system to study different usage scenarios.
Load in such systems is typically characterized by \textit{user limit} and \textit{spawn rate}.
User limit denotes the number of concurrent users accessing the system, while spawn rate denotes the rate at which new users are generated~\cite{ContainerizedGrouping}.
To introduce decent variety, we decided on three load levels: \texttt{low}, \texttt{medium}, and \texttt{high}.
Following \citet{ContainerizedGrouping} in using real website traffic for representative load testing, we extrapolated from the monthly traffic of popular websites reported by Semrush\footnote{\url{https://www.semrush.com}}.
For \texttt{high}, we used one-tenth of the number of visits of Amazon\footnote{\url{https://www.amazon.com}}, the most visited retail website in the world, i.e., 175 requests per second (the real number was too large for our experiment cluster).
We set the \texttt{low} level to the same number of visits as Bol\footnote{\url{https://www.bol.com}}, a local retail website in the Netherlands, i.e., 52 requests per second.
For \texttt{medium}, we then simply used the average of \texttt{low} and \texttt{high}, i.e., 114 request per second.
To nicely align these numbers with our experiment trial length of 60 seconds, we rounded the levels to 60, 120, and 180 respectively.

Regarding \textbf{dependent variables}, the most important one was energy consumption measured in joules (J).
For performance, we use two dependent variables, namely throughput, i.e., the number of requests the system handles for a specific time range and resource allocation, and response time, i.e., the duration until a client has received a response from the system.
Lastly, for maintainability, we used the five available sub-characteristics from \textit{Sigrid}:

\begin{itemize}
    \item Duplication: frequency of code clones in the system
    \item Unit Size: size of functions or methods in the system
    \item Unit Complexity: \# of decisions made in system units
    \item Unit Interfacing: complexity of unit interfaces
    \item Module Coupling: \# of dependencies a module has plus its size, with a module being a group of related units
\end{itemize}

Regarding hypotheses for testing, we expected each tactic or pattern to decrease energy consumption compared to the baseline, which led to the following null hypothesis:
\enquote{Implementing tactic T$_x$ or pattern P$_x$ in the system consumes the same or more energy as the baseline.}
Since it was not clear how the techniques would impact performance, we hypothesized that the variables would change, but not how exactly, leading to the following null hypothesis:
\enquote{Implementing tactic T$_x$ or pattern P$_x$ in the system results in the same response time / throughput as for the baseline.}
For the maintainability variables measured via \textit{Sigrid}, hypothesis testing was not suitable due to the benchmark-based nature of the analysis.
We therefore did not formulate any hypothesis, but simply compared the scores to the baseline to draw conclusions.

\subsection{Experiment Design and Execution}
\label{sec-Experiment Design}
Our experiment was based on 7 system versions (1 baseline, 3 tactics, 3 patterns) and 3 load levels (\texttt{low}, \texttt{medium}, \texttt{high}).
Using a full factorial design~\cite{Wohlin2024}, this resulted in 21 experiment configurations.
To avoid potential random confounders in the experiment environment, we repeated each configuration 30 times, leading to a total of 630 trials.
This was achieved by executing a randomized sequence of the 21 configurations 30 times, which had the additional benefit of minimizing potential infrastructure-level optimizations that might adapt to a fixed sequence.
Between each trial, we set a resting time of 1 minute so that the machine could cool down for more reliable energy measurements~\cite{cruz2021green}.
Before executing the experiment, we also turned off any unnecessary processes on the machine.

\begin{figure}[ht]
    \centering
    \includegraphics[width=0.85\linewidth]{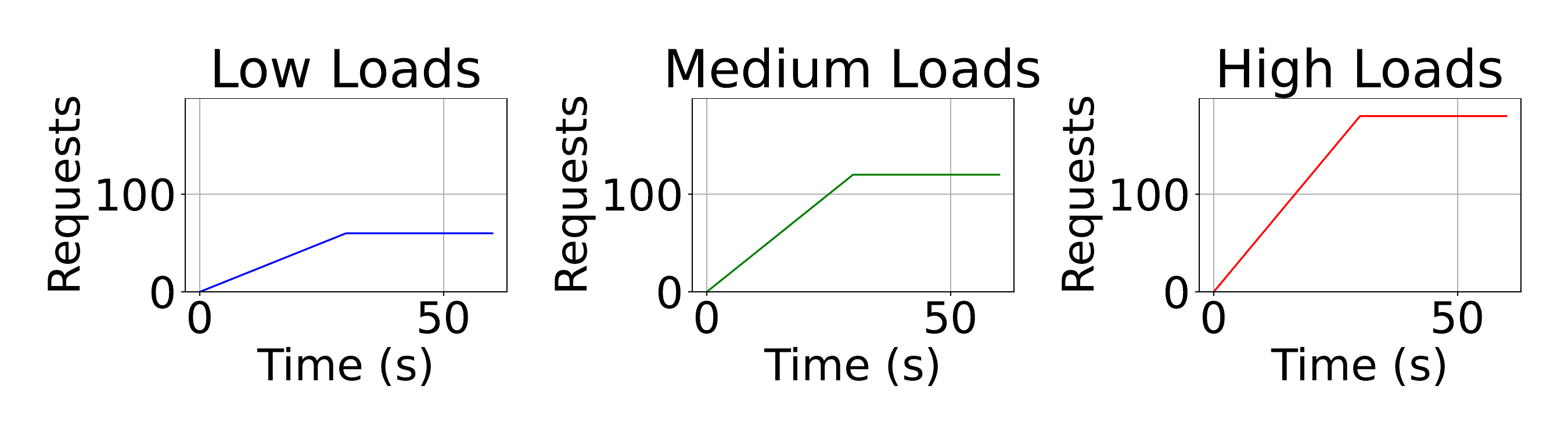}
    \caption{Load Generation for a 60-Second Experiment Trial}
    \label{fig:Users change under each load setting}
\end{figure}

During a trial, the load generator gradually increased the requests at a constant spawn rate over 30 seconds to reach the user limit, and then stayed at this user limit for another 30 seconds (see Fig.~\ref{fig:Users change under each load setting}).
This ensured that the system was not overloaded with a large number of requests without time to adjust its resource allocation.
For the analysis, we only considered the data after reaching the target user limit to observe the system's behavior under constant traffic, i.e., measurements were from the last 30 seconds of each trial.

\subsection{Data Analysis}
\label{s:Data Analysis}
After all experiment trials were finished, we collected the measurements from the respective tools and started with exploratory data analysis, e.g., by calculating descriptive statistics and creating visualizations like boxplots.
We also used the interquartile range method~\cite{vinutha2018detection} with the standard $1.5$ factor to detect and remove outliers.
Afterward, we analyzed if the data we wanted to use for hypothesis testing conformed to a normal distribution via the Shapiro-Wilk test~\cite{hanusz2016shapiro}.

Since all data of dependent variables turned out to be normally distributed (p-value $>0.05$), we selected the t-test as a suitable method~\cite{Student1908}.
Based on the described hypotheses, we used a one-tailed test for energy consumption and a two-tailed test for performance.
To combat the multiple comparisons problem due to testing many hypotheses~\cite{barnett2022multiple}, we applied the Holm-Bonferroni correction~\cite{holm1979simple} to adjust the corresponding p-value for each test.
For identified significant differences (adjusted p-value $< 0.05$), we used Cohen's $d$~\cite{cohen1988statistical} to estimate the effect size.
According to Cohen, values between $0.0$ and $0.2$ indicate a negligible effect, $0.2$ to $0.5$ a small effect, $0.5$ to $0.8$ a moderate effect, and $0.8$ and above a large effect.

For the maintainability-related metrics, we used the official continuous integration pipeline of \textit{Sigrid} to upload the codebase of each system version and then generated maintainability reports.
We then compared the scores for all metrics between the different versions.

\section{Results}
\label{s:results}
In this section, we present our results according to the RQs.
In tables, significant adjusted p-values are marked as bold and the effect sizes of insignificant tests are marked as \texttt{n/a}.

\subsection{Effectiveness of Reducing Energy Consumption (RQ1)}
\begin{figure}[ht]
    \centering
    \includegraphics[width=0.85\linewidth]{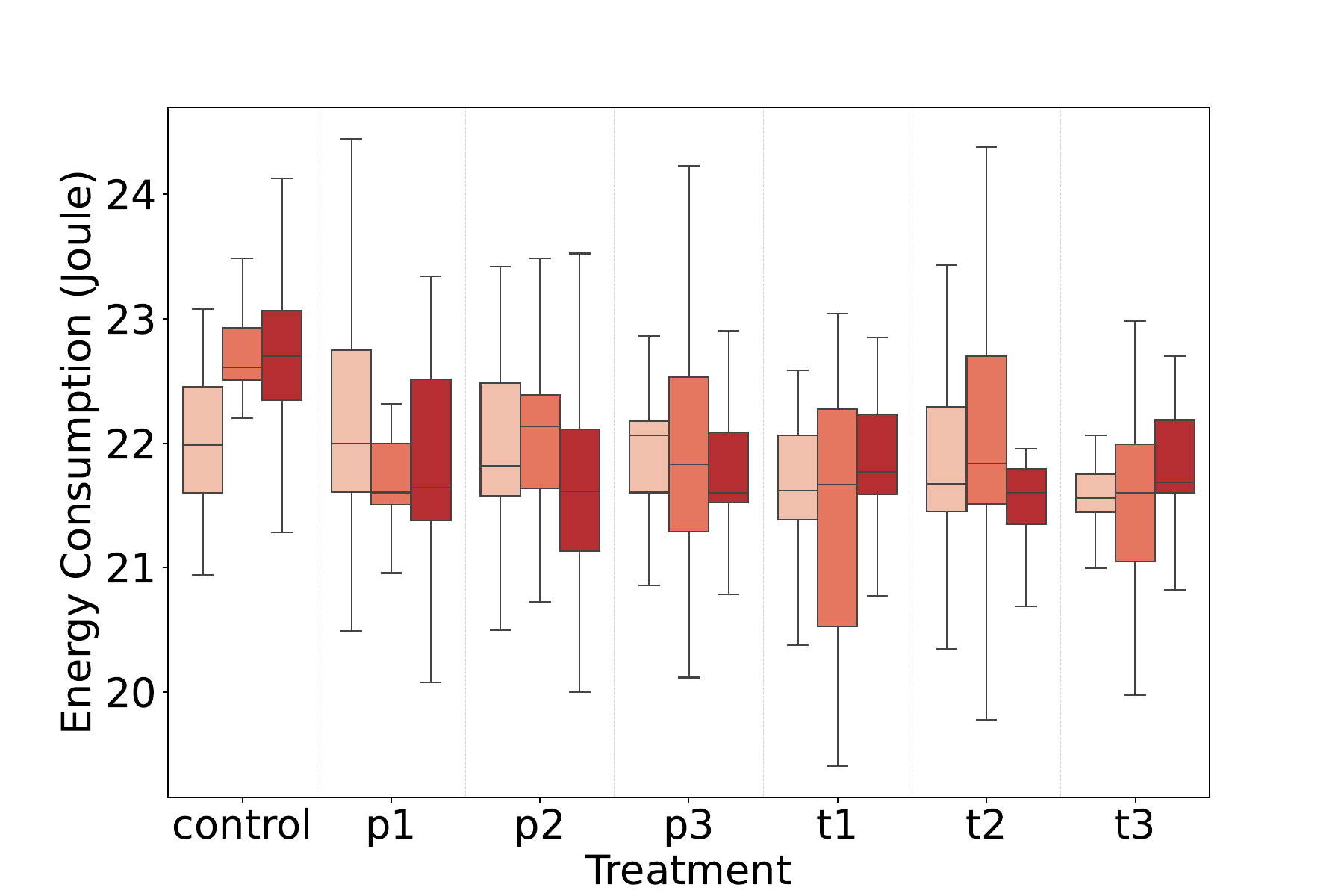}
    \\
    \footnotesize{\textbf{P1}: Backends for Frontends, \textbf{P2}: Request Bundle, \textbf{P3}: Caching of Read Requests, \textbf{T1}: Distribute pods to different nodes, \textbf{T2}: Distribute microservices, \textbf{T3}: Use different proxies,
    \\
    \textbf{Loads}: \colorbox{orange1}{LOW}, \colorbox{orange2}{\textcolor{white}{MEDIUM}}, \colorbox{orange3}{\textcolor{white}{HIGH}}}
    \caption{Energy Consumption Impact (J) of Techniques by Load}
    \label{fig:Energy Consumption by Tactics Patterns and Load}
\end{figure}

We show the energy consumption distribution of the techniques for each load level in Fig.~\ref{fig:Energy Consumption by Tactics Patterns and Load}.
The median values for the measured 1-second intervals of all treatments are in the range of 21 to 23 J, but with otherwise visibly different spreads.
We can also see that most techniques do not seem to reduce energy consumption when the load is low, but they all reduce energy consumption when the load is medium or high, some even considerably.
For medium load, \textit{Use different proxies} (T3) reduces the average energy consumption the most compared to the baseline (from 22.76 to 21.49 J, i.e., by 5.6\%), while for high load, it is instead \textit{Request Bundle} (P2) with the highest reduction (from 22.69 to 21.70 J, i.e., by 4.4\%).

To confirm if these mean differences were statistically significant, we performed t-tests and calculated Cohen's $d$ for significant differences to estimate the effect size.
We show the testing results, the effect sizes, and the mean percentage decrease in energy consumption in Table~\ref{tab:Energy Consumption Effect Size}.
As suspected from the visualization, nearly no technique significantly impacted energy consumption at low load.
The only exception was \textit{Use different proxies} (T3), which had a small effect (Cohen's $d$ of $-0.407$).
\textit{Distribute pods to different nodes} (T1) significantly reduced energy consumption for both medium load (large effect) and high load (moderate effect), indicating that load balancing can lead to a more energy-efficient system.
\textit{Distribute microservices} (T2) had a moderate effect on energy consumption for both medium and high load, with simplifying and reusing infrastructure contributing to efficiency.
In addition to its small effect for low load, \textit{Use different proxies} (T3) also significantly reduced energy consumption under medium load (large effect) and high load (moderate effect), suggesting that proxies play a vital role as load increases.
\textit{Backends for Frontends} (P1) showed a significant reduction for the higher load levels, with a large effect for medium and a moderate one for high loads, due to the benefits of fewer requests.
\textit{Request Bundle} (P2) showed a moderate effect for both medium and high loads, as the overhead of establishing and closing communications was reduced.
Finally, \textit{Caching of Read Requests} (P3) also significantly reduced energy consumption with a moderate effect for both medium and high loads, as fewer microservices needed to query databases and other services.

\begin{table}[ht]
    \centering
    \caption{Energy Consumption t-test Results and Effect Sizes}
    \label{tab:Energy Consumption Effect Size}
    \begin{tabular}{p{2.5cm}rrrr}
        \textbf{Techniques} & \textbf{Load} & \textbf{p-value} & \textbf{Cohen's $d$} & \textbf{Change} \\
        \hline
        \hline
        \multirow[t]{3}{=}{T1: Distribute pods to different nodes} & low & 0.13673 & n/a & n/a \\
        & medium & \textbf{$<$0.0001} & -1.153 & -5.5\% \\
        & high & \textbf{0.00038} & -0.648 & -3.6\% \\
        \multirow[t]{3}{=}{T2: Distribute microservices} & low & 0.28777 & n/a & n/a \\
        & medium & \textbf{0.00236} & -0.548 & -3.3\% \\
        & high & \textbf{0.00006} & -0.788 & -4.0\% \\
        \multirow[t]{3}{=}{T3: Use different proxies} & low & \textbf{0.01534} & -0.407 & -1.5\% \\
        & medium & \textbf{$<$0.0001} & -1.200 & -5.6\% \\
        & high & \textbf{0.00012} & -0.763 & -3.8\% \\
        \multirow[t]{3}{=}{P1: Backends for Frontends} & low & 0.90143 & n/a & n/a \\
        & medium & \textbf{$<$0.0001} & -1.961 & -4.4\% \\
        & high & \textbf{0.00038} & -0.686 & -3.8\% \\
        \multirow[t]{3}{=}{P2: Request Bundle} & low & 0.43587 & n/a & n/a \\
        & medium & \textbf{0.00340} & -0.543 & -2.6\% \\
        & high & \textbf{0.00008} & -0.775 & -4.4\% \\
        \multirow[t]{3}{=}{P3: Caching of Read Requests} & low & 0.49858 & n/a & n/a \\
        & medium & \textbf{0.00292} & -0.533 & -3.3\% \\
        & high & \textbf{0.00010} & -0.729 & -4.2\% \\
        \hline
        \hline
    \end{tabular}
\end{table}

\subsection{Impact on Performance (RQ2a)}
For performance, we analyzed the impact on two dependent variables, namely response time and throughput.

\subsubsection{Response Time}
\label{sec-Response Time}
\begin{figure}[ht]
    \centering
    \includegraphics[width=0.85\linewidth]{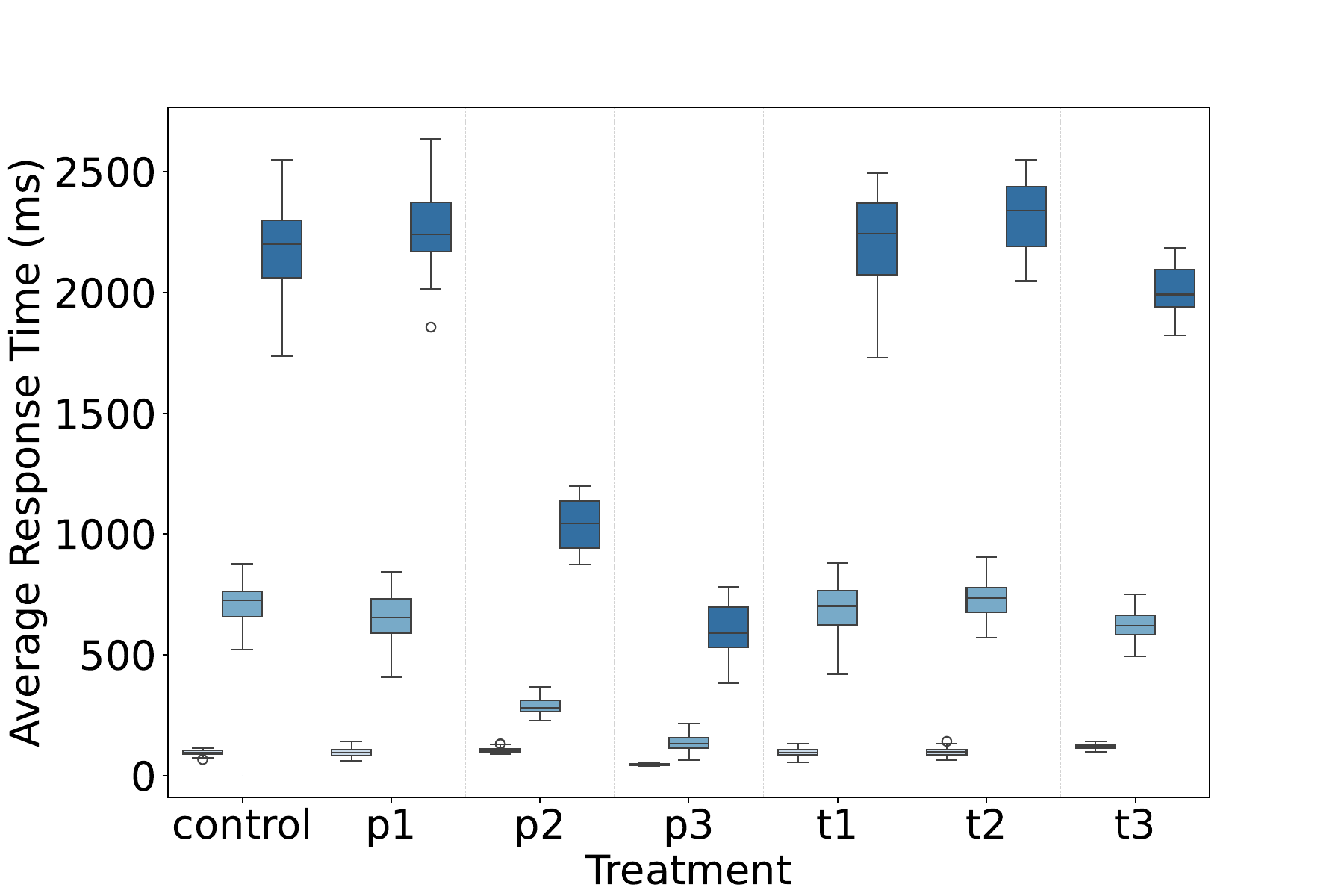}
    \\
    \footnotesize{\textbf{P1}: Backends for Frontends, \textbf{P2}: Request Bundle, \textbf{P3}: Caching of Read Requests, \textbf{T1}: Distribute pods to different nodes, \textbf{T2}: Distribute microservices, \textbf{T3}: Use different proxies, 
    \\
    \textbf{Loads}: \colorbox{blue1}{LOW}, \colorbox{blue2}{\textcolor{white}{MEDIUM}}, \colorbox{blue3}{\textcolor{white}{HIGH}}}
    \caption{Response Time Impact (ms) of Techniques by Load}
    \label{fig:Response Time by Tactics Patterns and Load}
\end{figure}

We show the response time distribution of each technique and load level in Fig.~\ref{fig:Response Time by Tactics Patterns and Load}.
This time, we see visibly larger differences between median values, even though they are not far apart for low load.
As expected, we also see consistently increasing response time with increasing load levels.
\textit{Caching of Read Requests} (P3) had the lowest mean response times across all load levels, which was especially notable under high load, with a mean of 599.6 ms.
\textit{Request Bundle} (P2) also showed a very low mean response time in medium and high loads compared to other tactics and patterns.
Other techniques did not seem to visibly reduce response time or even slightly increased it, e.g., the highest response times were achieved for \textit{Distribute microservices} (T2), with the mean being higher than the baseline in all cases, e.g., under high load, it was 2.32 s, an increase of 6.3\%.

We again continued with hypothesis testing and effect size estimation to confirm potential significant differences.
The results are shown in Table~\ref{tab:Response Time Effect Size}.
\textit{Distribute pods to different nodes} (T1) did not significantly affect response time at any load level.
This may seem counterintuitive, but the reason is simple: even the highest load level was consciously kept below a limit that would completely overload the system and cause it to fail responding.
Therefore, additional nodes were no advantage regarding response time.
Both \textit{Distribute microservices} (T2) and \textit{Backends for Frontends} (P1) only impacted response time significantly when the load was high.
In these cases, both techniques negatively affected the system by increasing response times (T2 with a moderate and P1 with a small effect).
\textit{Distribute microservices} (T2) compromised the independence of microservices, as several compatible services needed to be combined into a single container.
Some services in Online Boutique are programmed using single-threaded libraries or languages, and combining multiple such services limited their parallelism, which increased their response times.

The remaining techniques consistently had a significant impact on response time across all load levels, and mostly a positive one.
\textit{Use different proxies} (T3) showed a moderate effect for decreasing response time for medium and high load, but significantly increases response time when the load was low.
The impact of different proxies largely depends on the load scenario, but Istio seemed better suited for our load profiles.
\textit{Request Bundle} (P2) initially had a moderate negative impact on response time when the load was low.
However, with load increasing to medium and high, it instead substantially decreased response times with large effect sizes, making it one of our most effective techniques for performance.
When load was still low, the \textit{Request Bundle} response format carried redundant information, and the pattern required an additional element in the microservices to generate responses in an array-like way.
This overhead could only be offset once a certain load threshold was reached.
Lastly, \textit{Caching of Read Requests} (P3) was exceptionally effective at all load levels, significantly reducing response time with large effects, by far the strongest we saw for performance.

\begin{table}[ht]
    \centering
    \caption{Response Time t-test Results and Effect Sizes}
    \label{tab:Response Time Effect Size}
    \begin{tabular}{p{2.5cm}rrrr}
        \textbf{Techniques} & \textbf{Load} & \textbf{p-value} & \textbf{Cohen's $d$} & \textbf{Change} \\
        \hline
        \hline
        \multirow[t]{3}{=}{T1: Distribute pods to different nodes} 
        & low & 0.61676 & n/a & n/a \\
        & medium & 0.39465 & n/a & n/a \\
        & high & 0.47115 & n/a & n/a \\
        \multirow[t]{3}{=}{T2: Distribute microservices} 
        & low & 0.57048 & n/a & n/a \\
        & medium & 0.16260 & n/a & n/a \\
        & high & \textbf{0.00200} & 0.557 & 6.3\% \\
        \multirow[t]{3}{=}{T3: Use different proxies} 
        & low & \textbf{$<$0.0001} & 1.679 & 25.9\% \\
        & medium & \textbf{0.00022} & -0.665 & -11.5\% \\
        & high & \textbf{0.00007} & -0.748 & -7.8\% \\
        \multirow[t]{3}{=}{P1: Backends for Frontends} 
        & low & 0.78839 & n/a & n/a \\
        & medium & 0.06150 & n/a & n/a \\
        & high & \textbf{0.03878} & 0.358 & 3.6\% \\
        \multirow[t]{3}{=}{P2: Request Bundle} 
        & low & \textbf{0.00005} & 0.774 & 12.2\% \\
        & medium & \textbf{$<$0.0001} & -3.989 & -59.0\% \\
        & high & \textbf{$<$0.0001} & -4.955 & -52.3\% \\
        \multirow[t]{3}{=}{P3: Caching of Read Requests} 
        & low & \textbf{$<$0.0001} & -4.046 & -52.0\% \\
        & medium & \textbf{$<$0.0001} & -5.267 & -81.5\% \\
        & high & \textbf{$<$0.0001} & -7.338 & -72.5\% \\
        \hline
        \hline
    \end{tabular}
\end{table}

\subsubsection{Throughput}
\begin{figure}[ht]
    \centering
    \includegraphics[width=0.85\linewidth]{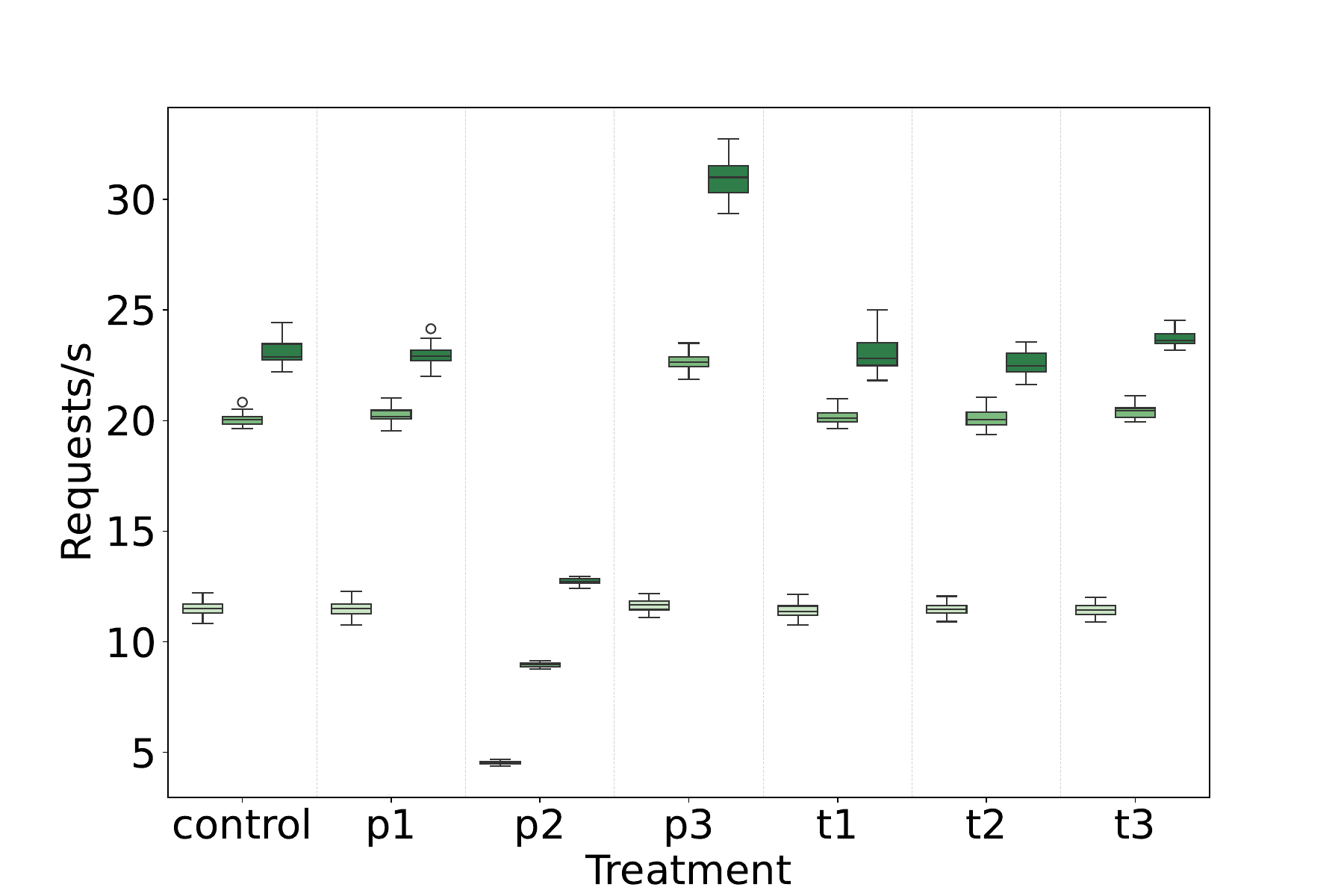}
    \\
    \footnotesize{\textbf{P1}: Backends for Frontends, \textbf{P2}: Request Bundle, \textbf{P3}: Caching of Read Requests, \textbf{T1}: Distribute pods to different nodes, \textbf{T2}: Distribute microservices, \textbf{T3}: Use different proxies, 
    \\
    \textbf{Loads}: \colorbox{green1}{LOW}, \colorbox{green2}{\textcolor{white}{MEDIUM}}, \colorbox{green3}{\textcolor{white}{HIGH}}}
    \caption{Throughput Impact (requests / s) of Techniques by Load}
    \label{fig:Requests per Second by Tactics Patterns and Load}
\end{figure}

We show the distribution of throughput per technique and load in Fig.~\ref{fig:Requests per Second by Tactics Patterns and Load}.
Throughput clearly increased with rising load levels, even though the jump from low to medium appears to have been mostly larger than the one from medium to high.
Many techniques visually seem to have performed similarly to the baseline, but we also see two noticeable differences.
First, \textit{Request Bundle} (P2) showed substantially lower throughput across all tested levels compared to other treatments and the control group.
However, due to how we measured throughput, these numbers are misleading and do not automatically indicate bad performance, which we will discuss further below.
Second, \textit{Caching of Read Requests} (P3) had the highest median throughput across all load levels, with great throughput especially for high load (30.9 requests / s).

To identify significant differences between treatments and baseline, we again conducted hypothesis testing and effect size estimation, for which we show the results in Table~\ref{tab:Throughput Effect Size}.
Except for \textit{Request Bundle} (P2) and \textit{Caching of Read Requests} (P3), no technique significantly impacted throughput at low load.
Moreover, as suspected from Fig.~\ref{fig:Requests per Second by Tactics Patterns and Load}, several techniques did not significantly affect throughput at all, or only for one load level.
As with response time, \textit{Distribute pods to different nodes} (T1) showed no significant effect, while \textit{Distribute microservices} (T2) negatively impacted throughput exclusively for high load (moderate effect).
Similarly, \textit{Backends for Frontends} (P1) generally had no significant impact, but showed a small positive effect on throughput when the load was medium.

Other patterns and tactics affected throughput more broadly.
While \textit{Using different proxies} (T3) had no effect for low load, it showed large positive effects when the load was medium and high.
However, the most effective technique to improve throughput was \textit{Caching of Read Requests} (P3), which showed a small effect for low load and then very large effects above that.
For example, it increased throughput at high loads when commonly used responses could be read from the cache more often by an astonishing average of 34\%.

Lastly, \textit{Request Bundle} (P2) performed most poorly for throughput on paper, with large negative effects at all load levels.
We defined throughput as the number of requests handled by the system within a certain time span, which is a standard definition used by many studies~\cite{de2016architecture, menasce2002load} as well as Locust.
Therefore, this reduction is caused by grouping multiple messages together, which naturally reduced the number of served requests in the measured timeframe.
In our implementation, \textit{Request Bundle} aggregated the messages of three operations in every request and response.
As a consequence, \textit{Request Bundle} would the best choice among all techniques if we changed the definition of throughput to the number of handled operations within a timeframe (three times the baseline).

\begin{table}[ht]
    \centering
    \caption{Throughput t-test Results and Effect Sizes}
    \label{tab:Throughput Effect Size}
    \begin{tabular}{p{2.5cm}rrrr}
        \textbf{Techniques} & \textbf{Load} & \textbf{p-value} & \textbf{Cohen's $d$} & \textbf{Change} \\
        \hline
        \hline
        \multirow[t]{3}{=}{T1: Distribute pods to different nodes} 
        & low & 0.07720 & n/a & n/a \\
        & medium & 0.09681 & n/a & n/a \\
        & high & 0.59181 & n/a & n/a \\
        \multirow[t]{3}{=}{T2: Distribute microservices} 
        & low & 0.47820 & n/a & n/a \\
        & medium & 0.68890 & n/a & n/a \\
        & high & \textbf{0.00023} & -0.684 & -2.3\% \\
        \multirow[t]{3}{=}{T3: Use different proxies} 
        & low & 0.24939 & n/a & n/a \\
        & medium & \textbf{$<$0.0001} & 0.993 & 1.9\% \\
        & high & \textbf{$<$0.0001} & 1.045 & 2.6\% \\
        \multirow[t]{3}{=}{P1: Backends for Frontends} 
        & low & 0.85943 & n/a & n/a \\
        & medium & \textbf{0.01028} & 0.445 & 1.1\% \\
        & high & 0.21324 & n/a & n/a \\
        \multirow[t]{3}{=}{P2: Request Bundle} 
        & low & \textbf{$<$0.0001} & -24.038 & -60.5\% \\
        & medium & \textbf{$<$0.0001} & -39.996 & -55.3\% \\
        & high & \textbf{$<$0.0001} & -19.423 & -44.9\% \\
        \multirow[t]{3}{=}{P3: Caching of Read Requests} 
        & low & \textbf{0.04575} & 0.340 & 1.1\% \\
        & medium & \textbf{$<$0.0001} & 5.230 & 13.1\% \\
        & high & \textbf{$<$0.0001} & 10.685 & 34.0\% \\
        \hline
        \hline
    \end{tabular}
\end{table}

\subsection{Impact on Maintainability (RQ2b)}
\label{sec:RQ2b}
Unlike for energy consumption and performance, the maintainability impact of used techniques is not affected by the different load levels, which we therefore could neglect for this analysis.
We show the maintainability sub-variable scores measured by \textit{Sigrid} in Table~\ref{tab:Maintainability Score - Unit Interfacing, Unit Size and Unit Complexity}.
Each score ranges from $0.5$ to $5.5$, with higher always being better.
These scores are based on a 5\% (0.5 - 1.5) -- 30\% (1.5 - 2.5) -- 30\% (2.5 - 3.5) -- 30\% (3.5 - 4.5) -- 5\% (4.5 - 5.5) distribution, which is drawn from the tool's past maintainability evaluations and comparisons of other software systems.
A higher score for each sub-variable is advantageous regarding development speed and costs.
Based on SIG's records, the development speed of a system with a score of $4$ is $3.5$ to $4$ times faster than for a system with a score of $2$, and the development speed of a system scored $5$ is $10$ times faster than for a system scored $1$.

\begin{table}[ht]
    \centering
    \caption{Maintainability Scores per Technique (UI: Unit Interfacing, US: Unit Size, UC: Unit Complexity, Dup: Duplication, MC: Module Coupling, scores from 0.5 to 5.5, higher is better)}
    \label{tab:Maintainability Score - Unit Interfacing, Unit Size and Unit Complexity}
    \begin{tabular}{lrrrrr}
        \textbf{Technique} & \textbf{UI} & \textbf{US} & \textbf{UC} & \textbf{Dup} & \textbf{MC} \\
        \hline
        \hline
        Control Group & 2.8 & 2.4 & 3.9 & 3.9 & 5.4 \\
        T1: Distribute pods to different nodes & 2.8 & 2.4 & 3.9 & 3.9 & 5.4 \\
        T2: Distribute microservices & 3.1 & 2.8 & 3.7 & 0.7 & 5.5 \\
        T3: Use different proxies & 2.8 & 2.4 & 3.9 & 3.9 & 5.4 \\
        P1: Backends for Frontends & 2.6 & 2.5 & 3.8 & 2.0 & 5.5 \\
        P2: Request Bundle & 2.7 & 2.8 & 4.1 & 4.0 & 5.4 \\
        P3: Caching of Read Requests & 2.9 & 2.6 & 3.4 & 4.0 & 2.9 \\
        \hline
        \hline
    \end{tabular}
\end{table}

\textit{Distributing pods to different nodes} (T1) did not change maintainability at the code level because it occurred exclusively in the Kubernetes Scheduler configuration, which was not analyzed by \textit{Sigrid}.
\textit{Distributing microservices} (T2) improved Unit Interfacing (+0.3) and Unit Size (+0.4), but then slightly worsened Unit Complexity (-0.2) and substantially worsened Duplication (-3.2). The latter was due to services with different dependencies being grouped, leading to the worst Duplication score of all techniques (0.7).
For example, some similar functions from two different microservices could not be merged because their dependent libraries were not the same or belonged to different non-compatible versions.
Overall, this tactic impacted maintainability negatively.
Like T1, \textit{Using different proxies} (T3) also did not affect code-level maintainability, as the changes were localized in the Kubernetes cluster configuration, not in individual microservices.
As \textit{Backends for Frontends} (P1) introduced new components and moved code around a lot, it had a varied impact across all sub-characteristics.
Some improved slightly through the pattern, namely Unit Size (+0.1) and Module Coupling (+0.1), while others like Unit Interfacing (-0.2) and Unit Complexity (-0.1) decreased slightly.
Most substantially, however, Duplication worsened considerably through the cloned functionality in the two backends (-1.9).
Overall, this pattern had a negative impact on maintainability.
\textit{Request Bundle} (P2) also had a fairly broad maintainability impact, with changes in 4 of the 5 characteristics.
While the request aggregating mechanism slightly harmed Unit Interfacing (-0.1), it improved Unit Size (+0.4), Unit Complexity (+0.2), and Duplication (+0.1).
This made \textit{Request Bundle} the only studied technique with an overall positive impact on maintainability.
Lastly, \textit{Caching of Read Requests} (P3) impacted all sub-characteristics.
While it slightly improved Unit Interfacing (+0.1), Unit Size (+0.2), and Duplication (+0.1), adding the decision-making logic and connecting it to the new in-memory cache module also moderately worsened Unit Complexity (-0.5) and strongly harmed Module Coupling (-2.5).
All in all, this pattern therefore still led to a negative impact on maintainability.

\begin{table*}[ht]
    \centering
    \caption{Summarized Increase (from $+$ to $+++$) or Decrease (from $-$ to $---$) in Dependent Variables Through Tactics and Patterns}
    \label{tab:Summary of Tactics and Patterns Effect Size on Quality Attributes}
    \begin{tabular}{llrrrr}
        \textbf{Technique} & \textbf{Load} & \textbf{Energy Consumption} & \textbf{Response Time} & \textbf{Throughput} & \textbf{Maintainability} \\
        \hline
        \hline
        \multirow[t]{3}{*}{T1: Distribute pods to different nodes} & Low    & n/a        & n/a        & n/a        & n/a        \\
                                         & Medium & $---$    & n/a        & n/a        & n/a  \\
                                         & High   & $--$ & n/a        & n/a        & n/a        \\
        \multirow[t]{3}{*}{T2: Distribute microservices} & Low    & n/a        & n/a        & n/a        & worse        \\
                                                  & Medium & $--$ & n/a        & n/a        & worse      \\
                                                  & High   & $---$    & $++$ & $--$ & worse        \\
        \multirow[t]{3}{*}{T3: Use different proxies} & Low    & $-$    & $---$    & n/a        & n/a        \\
                                               & Medium & $---$    & $--$ & $+++$    & n/a  \\
                                               & High   & $---$    & $--$ & $+++$    & n/a        \\
        \multirow[t]{3}{*}{P1: Backends for Frontends} & Low    & n/a        & n/a        & n/a        & worse        \\
                                                & Medium & $---$    & n/a        & $+$    & worse      \\
                                                & High   & $--$ & $+$    & n/a        & worse        \\
        \multirow[t]{3}{*}{P2: Request Bundle} & Low    & n/a        & $++$ & $---$    & better        \\
                                        & Medium & $--$ & $---$    & $---$    & better     \\
                                        & High   & $---$    & $---$    & $---$    & better        \\
        \multirow[t]{3}{*}{P3: Caching of Read Requests} & Low    & n/a        & $---$    & $+$    & worse        \\
                                             & Medium & $--$ & $---$    & $+++$    & worse      \\
                                             & High   & $---$    & $---$    & $+++$    & worse        \\
        \hline
        \hline
    \end{tabular}
\end{table*}

\section{Discussion}
\label{s:discussion}
In this section, we discuss important findings and the implications of our results.
Table~\ref{tab:Summary of Tactics and Patterns Effect Size on Quality Attributes} provides an aggregated summary of how strongly the tactics and patterns impacted our studied dependent variables, which can serve as a quick overview to identify the most useful techniques and potential trade-offs.
Note that $-$ and $+$ mean the decrease and increase in \textit{value}, not the nature of the impact.
For example, $---$ for the energy consumption or response time impact means a large decrease, which is beneficial.
Cells with \texttt{n/a} indicate no difference between the treatment and the control group.

One interesting finding of our experiment was that \textbf{most studied techniques did not significantly reduce energy consumption at low load}, except for \textit{Use different proxies} (T3), which had a small effect.
At this level, the energy saved through the techniques was likely equal or smaller to the energy overhead introduced by adding or enlarging components.
For example, with \textit{Distribute pods} (T1), the number of nodes with deployed microservices was higher than in the control group.
Under low load, the energy savings of load balancing between these nodes was insignificant, while the additional nodes increased the overall energy consumption.
Similarly, \textit{Backends for Frontends} (P1), \textit{Request Bundle} (P2), and \textit{Caching of Read Requests} (P3) all bring initial overheads for energy consumption if the new microservices and components they introduce into the system are not fully utilized, which is the case for low load.
This indicates that most studied techniques can only significantly and strongly reduce the energy consumption of microservices with consistently high request load.
Despite not being as pronounced as for energy consumption, we saw a similar relationship for performance.
For response time, 3 of 6 techniques had significant impact at low load, while only 2 of 6 techniques did so for throughput.

Another important result of our study was that \textbf{we can reduce the energy consumption of a microservice-based system without compromising its performance.}
For example, \textit{Caching of Read Requests} (P3) strongly improved both quality attributes, except for energy consumption at low load (no significant effect).
\textit{Using different proxies} (T3) also consistently improved both quality attributes for all load levels, but the strength of the effect varied.
Most energy was saved at higher loads with this tactic, but the response time improvement was the largest for low loads.
This highlights that, while it is possible to improve both energy consumption and response time simultaneously, the relationship between the two is not a simple linear one.
A last example in this category was \textit{Request Bundle} (P2), which was very load-sensitive.
At low load, it had no effect on energy consumption and even led to increased response times.
However, at medium and high load, the pattern was also able to substantially decrease energy consumption and response time, making it mostly suitable for systems that come under high request frequency.

However, some examined techniques were closer to a \textbf{trade-off relationship between energy consumption and response time}, e.g., \textit{Distribute microservices} (T2) and \textit{Backends for Frontends} (P1).
Both techniques had no significant impact at low load, but at higher load levels, they both managed to substantially reduce energy consumption.
At the highest load level, however, they simultaneously started to negatively impact response times.
This indicates that these techniques may not be suitable for saving energy at very high requests frequencies if trading off performance is not acceptable.

Furthermore, \textbf{several effective energy reduction techniques had negative impacts on maintainability}, namely \textit{Distribute microservices} (T2), \textit{Backends for Frontends} (P1), and \textit{Caching of Read Requests} (P3).
Especially under high loads, these techniques may substantially reduce energy consumption (in the case of P3 also response time), but they all add additional code or components, e.g., the caching logic, that make it more complex or time-consuming to maintain and evolve the system.
\textit{Distribute microservices} (T2) was one of the most challenging tactics to implement, as it involved solving compatibility problems residing in different versions of underlying libraries and rewriting many units.
It also violates one of the core principles of microservices architecture to have independently deployable services for increased flexibility and scalability.
Similar to performance, these maintainability trade-offs highlight that the current challenge is not simply reducing the energy consumption of microservices, as all six studied techniques were able to do this under high load.
Instead, the challenge is reducing energy consumption \textbf{while simultaneously not negatively impacting other QAs substantially}.

\section{Threats to Validity}
\label{sec:threats}
We discuss potential threats to the validity of our results based on the dimensions described by \citet{Wohlin2024}.
Regarding \textbf{construct validity}, one typical threat for controlled experiments is potentially unreliable or non-representative measurements of the dependent variables.
For example, we used Scaphandre to measure power at a specific time interval to derive energy consumption from it.
We did so once every second to have a decently small interval for accurate energy consumption measurements.
However, it should be noted that the result cannot be treated as continuous and smooth data.
Nonetheless, it is a common practice to approximate the energy consumption in this way, e.g., as is the case for the Running Average Power Limit (RAPL) algorithm that is widely adopted by Intel~\cite{hahnel2012measuring}.
We also tested Scaphandre extensively and observed that the power for the same treatment was very stable and reliable.
We did not use smaller time intervals (less than one second) because it might have led to a heavier burden on the experiment server.

Another threat is related to our maintainability measurements.
While the used maintainability sub-variables are based on ISO~25010 and \textit{Sigrid} has been widely evaluated in industry, our source-code-focused measurements did not reflect the whole scope of maintainability, especially given the architecture-centric nature of microservices.
As discussed in the results, the selected maintainability sub-variables could not detect some essential changes in our system, like the proxy configuration or node configuration change.
While reduced maintenance efforts are generally seen as a benefit of Istio~\cite{IstioImpact},
this was not reflected in the source code measurements.

Regarding \textbf{internal validity}, potential random events in the experiment environment can act as confounders to the results.
While our used experiment server is very professional, it does not have the same high availability as industrial cloud providers.
Its performance could sometimes fluctuate within a small range, and it sometimes also went completely offline.
Because a complete cycle of our experiment lasted several days, we experienced this interruption several times, and had to sometimes restart accordingly.
Our built experiment orchestrator partly mitigated the impact of this. 
Additionally, several researchers were using the experiment cluster in parallel.
While they were not using the same machines due to a shared calendar system, the simultaneous operations in the cluster could also have slightly impacted power or performance measurements.
Lastly, even something as simple as the room temperature can also affect the cluster performance and energy consumption measurements.
While such confounders can never be fully eliminated, we repeated each experiment configuration 30 times in randomized fashion, and therefore trust that such influence was negligible.

Regarding \textbf{external validity}, one potential threat is related to the implementation of patterns and tactics, as there is never a single way how to do this.
We aimed for simple and therefore less error-prone implementations, e.g., for Request Bundle, we
used a simple JSON structure to return multiple requests, while others may have chosen a more complex or XML-based structure.
There are also different frameworks and libraries that could be used for implementing the techniques.
To make our implementation more realistic, we always compared against the literature and examples in practitioner resources like cloud provider documentation, and discussed with our industry partner.
However, it is possible that a very different implementation would have led to different results.

Lastly, the single system we experimented with, Online Boutique, is an open-source e-commerce application, and naturally does not represent all types of microservice-based systems.
For example, it contains around 5k LOC distributed over 9 services plus one Web frontend, making it fairly small compared to many industry applications~\cite{Bogner2019}.
Despite many similarities, open-source projects cannot fully reflect all characteristics of closed-source industry software.
While we still believe in decent generalizability of our results, it is important to reproduce them across different systems and especially in real-world industry case studies.

\section{Conclusion}
\label{s:conclusion}
With this study, we have made progress in making microservice-based systems more energy-efficient by studying the effectiveness of proposed architectural tactics and patterns.
Several techniques can indeed reduce energy consumption while improving performance, which makes them especially valuable for industry.
While all techniques can significantly reduce energy consumption for higher loads, some techniques trade-off this gain against maintainability or performance.
We suggest that practitioners always assess system loads and their scaling before implementing tactics and patterns, as their effects are sensitive to load and application scenarios.

Since the energy efficiency of microservices is a fairly new topic, we anticipate that many new directions will emerge in the future.
For instance, future research should explore potential trade-offs of these techniques for different quality attributes or use other sustainability-related metrics, like water usage or carbon efficiency.
Additionally, studying the effects of combining several tactics and patterns in the same system will be important to understand the full potential of energy efficiency improvements.
To support such replications and extensions, we published our experiment artifacts on Zenodo.\footnote{\url{https://doi.org/10.5281/zenodo.12730144}}

\section*{Acknowledgment}
We kindly thank the Green Lab team at VU Amsterdam for providing and introducing us to the experiment infrastructure used in this study!

\bibliographystyle{IEEEtranN}
\bibliography{references}

\end{document}